# EQUIVALENT BOUNDARY CONDITIONS FOR THE ELECTROMAGNETIC FIELD ON THE SURFACE OF A PERIODIC GRATING OF RESISTIVE STRIPS

Igor M. Braver and Khona L. Garb

*Abstract*–The present work is aimed at formulating equivalent boundary conditions for the electromagnetic field on the surface of a periodic grating of resistive strips valid for arbitrary incident fields. It is assumed that the period of the grating is much smaller than the wavelength. We obtain two-sided boundary conditions of impedance type that allow one to replace any flat grating of resistive strips with a plane of discontinuity of the electromagnetic field.

## Introduction

Periodic gratings are widely used in the creation of polarizers, attenuators, interferometers, power dividers and other elements of microwave circuits [1, 2] operating in submillimeter range. Usually, period $l$ of periodic gratings (Fig. 1) is much smaller than the wavelength $\lambda$ in free space. In works [3-8] a quasi-static theory of diffraction of electromagnetic waves on periodic gratings is developed, which is valid under the condition $\frac{l}{\lambda} \ll 1$. A rigorous solution to the problem of diffraction on a flat grating of metal strips for arbitrary values of ratio $\frac{l}{\lambda}$ was obtained in the works [9, 10]. Extensive studies of the diffraction of electromagnetic waves on periodic gratings of various types were done by the Kharkiv School of Radiophysicists under the direction of Professor V. P. Shestopalov. The results of these studies are summarized in monographs [11-13].

In many works [3-8, 11, 14, 15] it has been shown that the influence of a fine $\left(\frac{l}{\lambda} \ll 1\right)$ periodic grating on the electromagnetic field can be described by introducing a plane of discontinuity of the electromagnetic field on which two-sided boundary conditions are satisfied. In the case of a flat grating of perfectly conducting strips, boundary conditions can be written in the form [14]:

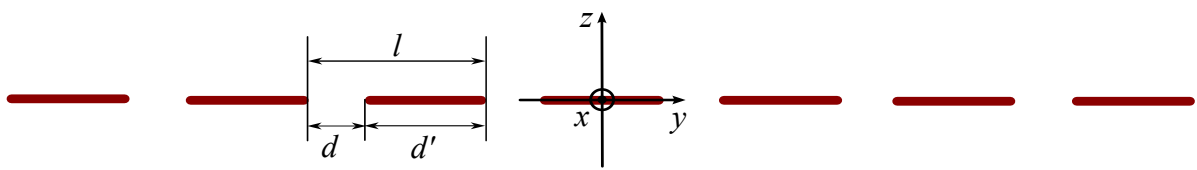

Fig. 1. Cross section of a flat periodic grating.

$$E_x\big|_{z=0} = -i\xi_1\left[\zeta_0\left(H_y\big|_{z=+0} - H_y\big|_{z=-0}\right) - \frac{i}{k_0}\frac{\partial}{\partial x}\left(E_z\big|_{z=+0} - E_z\big|_{z=-0}\right)\right],$$

$$H_x\big|_{z=+0} - H_x\big|_{z=-0} = -4i\xi_2\left[\eta_0 E_y\big|_{z=0} + \frac{i}{k_0}\frac{\partial}{\partial x}H_z\big|_{z=0}\right],$$

$$E_x\big|_{z=+0} = E_x\big|_{z=-0}, \qquad (1)$$

$$E_y\big|_{z=+0} = E_y\big|_{z=-0},$$

$$\xi_1 = \frac{k_0 l}{2\pi}\ln\left[\sin\frac{\pi d'}{2l}\right], \quad \xi_2 = \frac{k_0 l}{2\pi}\ln\left[\cos\frac{\pi d'}{2l}\right].$$

Here $k_0$ is the wave propagation constant in free space, $\zeta_0$ and $\eta_0$ are wave impedance and wave admittance of free space, **E** and **H** are the vectors of electric and magnetic fields with time dependence chosen in the form $\exp(-i\omega t)$.

Most works consider gratings consisting of perfectly conducting elements. The finite conductivity is taken into account in work [4], where equivalent boundary conditions are obtained for the case of a periodic grating of wires of round cross-section. In works [16] and [17], the Galerkin method was used to solve the problem of diffraction of E-polarized [16] and H-polarized [17] plane waves on a periodic grating of resistive strips under the assumption that the incident field does not depend on the *x*-coordinate along the strips. The aim of this work is to formulate equivalent boundary conditions for the electromagnetic field on the surface of a periodic grating of resistive strips valid for arbitrary, possibly *x*-dependent, incident fields.

## 1. Problem statement

Let the flat grating formed by resistive parallel strips lie in the *xy*-plane (Fig. 1). In the direction of the *y*-axis the grating is periodic with a period *l*. A linearly polarized plane wave, which is a superposition of H-polarized ($E_x \equiv 0$) and E-polarized $\left(H_x \equiv 0\right)$ waves, is incident onto the grating from the upper half-space:

$$E_x^0 = E_0 e^{-i(k_z z - k_y y)} e^{ik_x x},$$
$$H_x^0 = H_0 e^{-i(k_z z - k_y y)} e^{ik_x x}, \qquad (2)$$
$$k_x^2 + k_y^2 + k_z^2 = k_0^2.$$

Here $E_0$, $H_0$ are the amplitudes of the incident waves, $k_x$, $k_y$, $k_z$ are the components of the wave vector along the corresponding axis. In the case of perfectly conducting strips, the solution is represented by a superposition of two independent solutions for E-polarized and H- polarized waves [13]. This conclusion does not apply to resistive strips: the incidence of an E-polarized wave causes the appearance of both E- and H-polarized waves; accordingly, the incident H-polarized wave excites not only H- but also E-polarized waves.

According to Maxwell's equations, the electromagnetic field in the structure under consideration can be expressed through the longitudinal $E_x$ and $H_x$ components of the field:

$$E_y = \frac{i}{k_0^2 - k_x^2}\left(k_x \frac{\partial E_x}{\partial y} + k_0 \zeta_0 \frac{\partial H_x}{\partial z}\right),$$

$$E_z = \frac{i}{k_0^2 - k_x^2}\left(k_x \frac{\partial E_x}{\partial z} - k_0 \zeta_0 \frac{\partial H_x}{\partial y}\right),$$

$$H_y = \frac{-i}{k_0^2 - k_x^2}\left(k_0 \eta_0 \frac{\partial E_x}{\partial z} - k_x \frac{\partial H_x}{\partial y}\right), \quad (3)$$

$$H_z = \frac{i}{k_0^2 - k_x^2}\left(k_0 \eta_0 \frac{\partial E_x}{\partial y} + k_x \frac{\partial H_x}{\partial z}\right).$$

Since the structure under consideration is uniform in the direction of the $x$ axis, the dependence of the total field on the $x$ coordinate will be the same as in the incident wave. Therefore, the components $E_x$ and $H_x$ satisfy the two-dimensional Helmholtz equation

$$\nabla^2 E_x + \left(k_0^2 - k_x^2\right) E_x = 0,$$

$$\nabla^2 H_x + \left(k_0^2 - k_x^2\right) H_x = 0, \quad (4)$$

$$\nabla^2 = \frac{\partial^2}{\partial y^2} + \frac{\partial^2}{\partial z^2}.$$

The components of the electromagnetic field tangential to the grating are continuous at the slits, and at the resistive strips they satisfy the known two-sided boundary conditions [18]. Their essence lies in the fact that the tangential components of the electric field are continuous when moving from one side of the resistive surface to the other, while the tangential components of the magnetic field experience a jump whose magnitude is proportional to the surface admittance $Y$ of the resistive film. Thus, the boundary conditions for the electromagnetic field in the plane $z = 0$ of the periodic grating can be written in the following form

$$\left.\begin{array}{l} E_x\big|_{z=+0} = E_x\big|_{z=-0}, \\ E_y\big|_{z=+0} = E_y\big|_{z=-0}, \end{array}\right\} \quad -\frac{l}{2} + nl \leq y \leq \frac{l}{2} + nl, \quad (5a)$$

$$\left.\begin{array}{l} E_x\big|_{z=0} = -W\left(H_y\big|_{z=+0} - H_y\big|_{z=-0}\right), \\ E_y\big|_{z=0} = W\left(H_x\big|_{z=+0} - H_x\big|_{z=-0}\right), \end{array}\right\} \quad -\frac{d'}{2} + nl \leq y \leq \frac{d'}{2} + nl. \quad (5b)$$

Here $n = 0, \pm 1, \pm 2, \ldots$, and $W = \dfrac{1}{Y}$ is the surface impedance of the resistive film.

The electromagnetic field is a periodic function of the $y$ coordinate. Therefore $E_x$- and $H_x$- components of the field scattered by the grating should be sought for in the form of a Fourier series

$$E_x(y,z) = \sum_{n=-\infty}^{\infty} E_n(z) e^{i\left(k_y + \frac{2\pi n}{l}\right)y},$$
$$H_x(y,z) = \sum_{n=-\infty}^{\infty} H_n(z) e^{i\left(k_y + \frac{2\pi n}{l}\right)y}. \tag{6}$$

Hereafter we omit the factor $\exp(ik_x x)$ which describes the dependence of the field on the $x$ coordinate.

Substituting $E_x$ and $H_x$ from (6) into the Helmholtz equation (4), we find that $E_n(x)$ and $H_n(x)$ satisfy the differential equation

$$\left(\frac{\partial^2}{\partial z^2} + \gamma_n^2\right) \begin{Bmatrix} E_n(z) \\ H_n(z) \end{Bmatrix} = 0,$$
$$\gamma_n^2 = k_0^2 - k_x^2 - h_n^2, \quad h_n = k_y + \frac{2\pi n}{l}. \tag{7}$$

Note that $\gamma_0$ and $h_0$ coincide with the $z$- and $y$-components of the incident wave vector:

$$\gamma_0 = k_z, \quad h_0 = k_y. \tag{8}$$

The general solution of the differential equation (7) has the form

$$E_n(z) = a_n \exp(i\gamma_n z) + b_n \exp(-i\gamma_n z),$$
$$H_n(z) = A_n \exp(i\gamma_n z) + B_n \exp(-i\gamma_n z), \tag{9}$$

where $a_n$, $b_n$, $A_n$, $B_n$ are arbitrary constants, and terms with factors $\exp(i\gamma_n z)$ [$\exp(-i\gamma_n z)$] describe waves propagating in the positive [negative] direction of the $z$-axis.

Thus, the $E_x$- and $H_x$-components of the total field for the structure under consideration can be written in the following form

$$E_x(y,z) = E_0 e^{-ik_z z} e^{ik_y y} + \sum_{n=-\infty}^{\infty} a_n e^{i\gamma_n z} e^{ih_n y}, \quad z > 0,$$
$$E_x(y,z) = \sum_{n=-\infty}^{\infty} b_n e^{-i\gamma_n z} e^{ih_n y}, \quad z < 0. \tag{10}$$

$$H_x(y,z) = H_0 e^{-ik_z z} e^{ik_y y} + \sum_{n=-\infty}^{\infty} A_n e^{i\gamma_n z} e^{ih_n y}, \quad z > 0,$$
$$H_x(y,z) = \sum_{n=-\infty}^{\infty} B_n e^{-i\gamma_n z} e^{ih_n y}, \quad z < 0. \tag{11}$$

Components $E_y$ and $H_y$ can be expressed using (3) in terms of the unknown coefficients $a_n, b_n; A_n, B_n$:

$$E_y(y,z) = \frac{-1}{k_0^2 - k_x^2} \begin{pmatrix} k_x k_y E_0 e^{-ik_z z} e^{ik_y y} - H_0 \varsigma_0 k_0 k_z e^{-ik_z z} e^{ik_y y} + \\ + k_x \sum_{n=-\infty}^{\infty} h_n a_n e^{i\gamma_n z} e^{ih_n y} + k_0 \varsigma_0 \sum_{n=-\infty}^{\infty} \gamma_n A_n e^{i\gamma_n z} e^{ih_n y} \end{pmatrix}, \quad z > 0, \tag{12a}$$

$$E_y(y,z) = \frac{-1}{k_0^2 - k_x^2}\left(k_x \sum_{n=-\infty}^{\infty} h_n b_n e^{-i\gamma_n z} e^{ih_n y} - k_0 \zeta_0 \sum_{n=-\infty}^{\infty} \gamma_n B_n e^{-i\gamma_n z} e^{ih_n y}\right), \quad z < 0, \qquad (12b)$$

$$H_y(y,z) = \frac{-1}{k_0^2 - k_x^2}\begin{pmatrix} \eta_0 k_0 k_z E_0 e^{-ik_z z} e^{ik_y y} + k_x k_y H_0 e^{-ik_z z} e^{ik_y y} - \\ -k_0 \eta_0 \sum_{n=-\infty}^{\infty} \gamma_n a_n e^{i\gamma_n z} e^{ih_n y} + k_x \sum_{n=-\infty}^{\infty} h_n A_n e^{i\gamma_n z} e^{ih_n y} \end{pmatrix}, \quad z > 0, \qquad (13a)$$

$$H_y(y,z) = \frac{-1}{k_0^2 - k_x^2}\left(k_0 \eta_0 \sum_{n=-\infty}^{\infty} \gamma_n b_n e^{-i\gamma_n z} e^{ih_n y} + k_x \sum_{n=-\infty}^{\infty} h_n B_n e^{-i\gamma_n z} e^{ih_n y}\right), \quad z < 0. \qquad (13b)$$

From the continuity of the $E_x$- and $E_y$-components of the electric field in the plane $z = 0$ of the grating it follows that

$$\begin{aligned} E_0 + a_0 &= b_0, \quad a_n = b_n, \\ -H_0 + A_0 &= -B_0, \quad A_n = -B_n. \end{aligned} \qquad (14)$$

The next step is to find the coefficients $b_n$ and $B_n$. In works [10-13, 15], this problem for the case of perfectly conducting strips was approached by formulating a system of functional equations based on the boundary conditions for the electromagnetic field. This system of functional equations was solved by the Riemann-Hilbert method. Using the approximation $k_0 l \ll 1$, explicit expressions were obtained for the coefficients $b_0$ and $B_0$, allowing one to formulate the equivalent boundary conditions for the electromagnetic field. In the case of perfectly conducting strips, it is sufficient to solve the problem for one polarization, since, in accordance with the duality principle, the solution for the other polarization is obtained from the solution for the first one if the slits and strips are interchanged. In the case of resistive strips, it is impossible to find a rigorous solution. Therefore, we will use a different approach. Based on the two-sided boundary conditions (5b), we formulate two vector integral equations (IE): one for the jump of the tangential magnetic field on the film, and another one for the tangential electric field on the interval $\frac{-l}{2} \leq y \leq \frac{l}{2}$. Using these equations, we formulate stationary expressions for the elements of the scattering matrix, from which we derive the desired equivalent conditions in the static approximation $k_0 l \ll 1$.

## 2. Vector integral equation for the magnetic field jump on a film

### 2.1 General case

Let us find the jump of the tangential components of the magnetic field on the film:

$$\begin{aligned} H_x\big|_{z=+0} - H_x\big|_{z=-0} &= 2\left[H_0 e^{ik_y y} - \sum_{n=-\infty}^{\infty} B_n e^{ih_n y}\right], \\ H_y\big|_{z=+0} - H_y\big|_{z=-0} &= \frac{-2}{k_0^2 - k_x^2}\begin{bmatrix} \eta_0 k_0 k_z E_0 e^{ik_y y} + k_x k_y H_0 e^{ik_y y} - \\ -\eta_0 k_0 \sum_{n=-\infty}^{\infty} \gamma_n b_n e^{ih_n y} - k_x \sum_{n=-\infty}^{\infty} h_n B_n e^{ih_n y} \end{bmatrix}. \end{aligned} \qquad (15)$$

Let us denote

$$H_x\big|_{z=+0} - H_x\big|_{z=-0} = h_x(y)e^{ik_y y},$$
$$H_y\big|_{z=+0} - H_y\big|_{z=-0} = h_y(y)e^{ik_y y}. \qquad (16)$$

Then

$$H_0 - \sum_{n=-\infty}^{\infty} B_n e^{i\frac{2\pi n}{l}y} = \frac{1}{2}h_x(y). \qquad (17)$$

Taking into account that outside the resistive strip $\mathbf{h}(y) \equiv 0$, from (17) based on Fourier's theorem we find

$$B_n - \delta_{n0}H_0 = -A_n = -\frac{1}{2l}\int_{-\frac{d'}{2}}^{\frac{d'}{2}} dy\, h_x(y) e^{-i\frac{2\pi n}{l}y}, \quad n=0,\pm 1,\pm 2,\ldots \qquad (18)$$

Similarly, using the second expression from (15) and formula (18), we obtain integral representations of the coefficients $a_n$, $b_n$:

$$b_n - \delta_{n0}E_0 = a_n =$$
$$= \zeta_0 \frac{k_0^2 - k_x^2}{2lk_0\gamma_n} \int_{-\frac{d'}{2}}^{\frac{d'}{2}} dy\, h_y(y) e^{-i\frac{2\pi n}{l}y} + \zeta_0 \frac{k_x h_n}{2lk_0\gamma_n} \int_{-\frac{d'}{2}}^{\frac{d'}{2}} dy\, h_x(y) e^{-i\frac{2\pi n}{l}y}. \qquad (19)$$

Let us now express the electric field in the entire space in terms of the jump $\mathbf{h}(y)$ of the magnetic field on the film

$$E_x(y,z) = E_0 e^{-ik_z z} e^{ik_y y} +$$
$$+ e^{ik_y y}\int_{-\frac{d'}{2}}^{\frac{d'}{2}} dy'\left[\sum_{n=-\infty}^{\infty} \zeta_0 \frac{k_x h_n}{2k_0 l\gamma_n} e^{i\frac{2\pi n}{l}(y-y')} e^{i\gamma_n|z|}\right] h_x(y') + \qquad (20)$$
$$+ e^{ik_y y}\int_{-\frac{d'}{2}}^{\frac{d'}{2}} dy'\left[\sum_{n=-\infty}^{\infty} \zeta_0 \frac{k_0^2 - k_x^2}{2k_0 l\gamma_n} e^{i\frac{2\pi n}{l}(y-y')} e^{i\gamma_n|z|}\right] h_y(y'),$$

$$E_y(y,z) = \frac{1}{k_0^2 - k_x^2}\left(-k_x k_y E_0 + k_0 k_z \zeta_0 H_0\right) e^{-ik_z z} e^{ik_y y} -$$
$$- e^{ik_y y}\int_{-\frac{d'}{2}}^{\frac{d'}{2}} dy'\left[\sum_{n=-\infty}^{\infty} \zeta_0 \frac{k_0^2 - h_n^2}{2k_0 l\gamma_n} e^{i\frac{2\pi n}{l}(y-y')} e^{i\gamma_n|z|}\right] h_x(y') - \qquad (21)$$
$$- e^{ik_y y}\int_{-\frac{d'}{2}}^{\frac{d'}{2}} dy'\left[\sum_{n=-\infty}^{\infty} \zeta_0 \frac{k_x h_n}{2k_0 l\gamma_n} e^{i\frac{2\pi n}{l}(y-y')} e^{i\gamma_n|z|}\right] h_y(y').$$

Substituting expressions (20), (21) into the two-sided boundary conditions (5b), we obtain a vector equation for determining $\mathbf{h}(y)$:

$$\int_{-\frac{d'}{2}}^{\frac{d'}{2}} dy' \zeta_{xx}(y, y') h_x(y') + \int_{-\frac{d'}{2}}^{\frac{d'}{2}} dy' \zeta_{xy}(y, y') h_y(y') + W h_x(y) =$$

$$= -\frac{k_x k_y}{k_0^2 - k_x^2} E_0 + \frac{k_0 k_z}{k_0^2 - k_x^2} \zeta_0 H_0, \tag{22}$$

$$-\frac{d'}{2} \leq y \leq \frac{d'}{2},$$

$$\int_{-\frac{d'}{2}}^{\frac{d'}{2}} dy' \zeta_{yx}(y, y') h_x(y') + \int_{-\frac{d'}{2}}^{\frac{d'}{2}} dy' \zeta_{yy}(y, y') h_y(y') + W h_y(y) = -E_0,$$

where the components of the tensor $\zeta(y, y')$ are given by

$$\zeta_{xx}(y, y') = \frac{\zeta_0}{2k_0 l} \sum_{n=-\infty}^{\infty} \frac{k_0^2 - h_n^2}{\gamma_n} e^{i\frac{2\pi n}{l}(y-y')},$$

$$\zeta_{xy}(y, y') = \frac{\zeta_0 k_x}{2k_0 l} \sum_{n=-\infty}^{\infty} \frac{h_n}{\gamma_n} e^{i\frac{2\pi n}{l}(y-y')},$$

$$\zeta_{yx}(y, y') = \zeta_{xy}(y, y'), \tag{23}$$

$$\zeta_{yy}(y, y') = \frac{\zeta_0 (k_0^2 - k_x^2)}{2k_0 l} \sum_{n=-\infty}^{\infty} \frac{1}{\gamma_n} e^{i\frac{2\pi n}{l}(y-y')}.$$

### 2.2. Perfectly conducting strips

Let us consider separately the case when the film is perfectly conducting ($W = 0$). As is known [13], in this case the problem is divided into two independent problems:

a) For an E-polarized wave, when $H_0 = 0$, $H_x(y, z) = 0$, $h_x(y) = 0$, and all field components $(H_y, H_z, E_y, E_z)$ are expressed in terms of $E_x$ (see formulas (3) for $H_x = 0$).

b) For an H-polarized wave, when $E_0 = 0$, $E_x(y, z) = 0$, and components $E_y, E_z, H_y, H_z$ are expressed in terms of $H_x$.

In the case of an E-polarized wave, substituting $H_0 = 0$, $h_x(y) = 0$ into the IE (22), we find

$$\int_{-\frac{d'}{2}}^{\frac{d'}{2}} dy' \zeta_{xy}(y, y') h_y(y') = -\frac{k_x k_y}{k_0^2 - k_x^2} E_0,$$

$$\int_{-\frac{d'}{2}}^{\frac{d'}{2}} dy' \zeta_{yy}(y, y') h_y(y') = -E_0. \tag{24}$$

Let us show that the solution of the second IE from system (24) also satisfies the first IE. To prove this, we multiply the second IE by $\exp(ik_y y)$. Then, we differentiate the resulting equality with respect to $y$ and take into account that $k_y + \dfrac{2\pi n}{l} = h_n$. This way we arrive at the first IE.

In the case of an H-polarized wave, the $H_y$- component, as follows from (3), is expressed in terms of the $H_x$-component:

$$H_y(y,z) = \frac{ik_x}{k_0^2 - k_x^2} \frac{\partial H_x(y,z)}{\partial y}. \tag{25}$$

From here, taking into account (16), we find

$$h_y(y) = e^{-ik_y y} \frac{ik_x}{k_0^2 - k_x^2} \frac{\partial}{\partial y}\left(h_x e^{ik_y y}\right). \tag{26}$$

Substituting (26) into the system of equations (22), where $W = 0$ is assumed, integrating by parts and taking into account [19, 20] that $h_x\left(\pm\dfrac{d'}{2}\right) = 0$, we obtain that the second equation from the system (22) is satisfied identically, while the first equation is transformed to the following form

$$\frac{\zeta_0}{2k_0 l} \int_{-\frac{d'}{2}}^{\frac{d'}{2}} dy' \left[\sum_{n=-\infty}^{\infty} \gamma_n e^{i\frac{2\pi n}{l}(y-y')}\right] h_x(y') = \zeta_0 \frac{k_z}{k_0} H_0. \tag{27}$$

2.3 Oblique incidence of electromagnetic wave $\left(\dfrac{\partial}{\partial x} = 0\right)$

Let us also consider the case, when the field of the incident wave does not depend on $x$ coordinate, i.e. $\dfrac{\partial}{\partial x} = ik_x = 0$. In the monograph [13], such a case is called the case of oblique incidence, unlike the case of arbitrary incidence, when $k_x \neq 0$. At $k_x = 0$, elements (23) of the impedance tensor $\zeta$ take the following form

$$\begin{aligned}
\zeta_{xx}(y, y') &= \frac{\zeta_0}{2k_0 l} \sum_{n=-\infty}^{\infty} \gamma_n e^{i\frac{2\pi n}{l}(y-y')}, \\
\zeta_{xy}(y, y') &= \zeta_{yx}(y, y') = 0, \\
\zeta_{yy}(y, y') &= \frac{\zeta_0 k_0}{2l} \sum_{n=-\infty}^{\infty} \frac{e^{i\frac{2\pi n}{l}(y-y')}}{\gamma_n}.
\end{aligned} \tag{28}$$

Substituting (28) into the system of IE (22), we confirm that for oblique incidence of the wave $(k_x = 0)$, the vector diffraction problem is divided into two scalar problems.

a) Scattering of an H-polarized wave. In this case, the electromagnetic field has components $H_x$, $E_y$, $E_z$ and the only component $h_x$ of the magnetic field jump satisfies the IE

$$\int_{-\frac{d'}{2}}^{\frac{d'}{2}} dy' \zeta_{xx}(y, y') h_x(y') + W h_x(y) = \zeta_0 \frac{k_z}{k_0} H_0. \tag{29}$$

b) Incidence of an E-polarized wave. In this case, the electromagnetic field has components $E_x$, $H_y$, $H_z$. The only component $h_y$ of the magnetic field jump satisfies the IE

$$\int_{-\frac{d'}{2}}^{\frac{d'}{2}} dy' \zeta_{yy}(y, y') h_y(y') + W h_y(y) = -E_0. \tag{30}$$

In the case of normal incidence not only $k_x = 0$, but also $k_y = 0$. Then

$$k_z = k_0, \ h_n = \frac{2\pi n}{l}, \ \gamma_n^2 = k_0^2 - h_n^2, \ \gamma_n = \gamma_{-n}. \tag{31}$$

Let us consider in some more detail the problem of normal incidence of an E-polarized wave. Taking into account the equality $\gamma_n = \gamma_{-n}$, the kernel of the IE (30) can be transformed to the form

$$\zeta_{yy}(y, y') = \zeta_0 \frac{k_o}{2l} \sum_{n=0}^{\infty} \varepsilon_n \frac{\cos\left(\frac{2\pi n}{l} y\right) \cos\left(\frac{2\pi n}{l} y'\right) + \sin\left(\frac{2\pi n}{l} y\right) \sin\left(\frac{2\pi n}{l} y'\right)}{\gamma_n}. \tag{32}$$

As can be seen from (32), $\zeta_{yy}$ can be divided into two parts, one of which is an even function of $y, y'$, and the other is an odd function of the same variables:

$$\zeta_{yy}(y, y') = \zeta_{yy}^e(y, y') + \zeta_{yy}^o(y, y'),$$
$$\zeta_{yy}^e(y, y') = \zeta_{yy}^e(y, -y') = \zeta_{yy}^e(-y, y'), \tag{33}$$
$$\zeta_{yy}^o(y, y') = -\zeta_{yy}^o(y, -y') = -\zeta_{yy}^o(-y, y').$$

Accordingly, we will look for the solution of the integral equation (30) in the form of a sum of two terms, one of which is an even and the other an odd function of the argument $y$:

$$h_y(y) = h_y^e(y) + h_y^o(y),$$
$$h_y^e(y) = h_y^e(-y), \ h_y^o(y) = -h_y^o(-y). \tag{34}$$

Let us write the equation for two points symmetrical with respect to the origin:

$$\left[\int_{-\frac{d'}{2}}^{\frac{d'}{2}} dy' \zeta_{yy}^{e}(y,y') h_{y}^{e}(y') + W h_{y}^{e}(y)\right] + \left[\int_{-\frac{d'}{2}}^{\frac{d'}{2}} dy' \zeta_{yy}^{o}(y,y') h_{y}^{o}(y') + W h_{y}^{o}(y)\right] = -E_0,$$

$$\left[\int_{-\frac{d'}{2}}^{\frac{d'}{2}} dy' \zeta_{yy}^{e}(-y,y') h_{y}^{e}(y') + W h_{y}^{e}(-y)\right] + \left[\int_{-\frac{d'}{2}}^{\frac{d'}{2}} dy' \zeta_{yy}^{o}(-y,y') h_{y}^{o}(y') + W h_{y}^{o}(-y)\right] = -E_0,$$

(35)

Summing and subtracting these equations and taking into account the parity properties (33), (34), we obtain

$$\int_{-\frac{d'}{2}}^{\frac{d'}{2}} dy' \zeta_{yy}^{e}(y,y') h_{y}^{e}(y') + W h_{y}^{e}(y) = -E_0, \tag{36a}$$

$$\int_{-\frac{d'}{2}}^{\frac{d'}{2}} dy' \zeta_{yy}^{o}(y,y') h_{y}^{o}(y') + W h_{y}^{o}(y) = 0. \tag{36b}$$

From (36b) it follows that

$$h_{y}^{o}(y) = 0. \tag{37}$$

Equation (36a) completely coincides with the integral equation for the problem of scattering of the lowest H-type wave ($H_{00}$) on a symmetric resistive film in a plane waveguide with magnetic walls (Fig. 2)

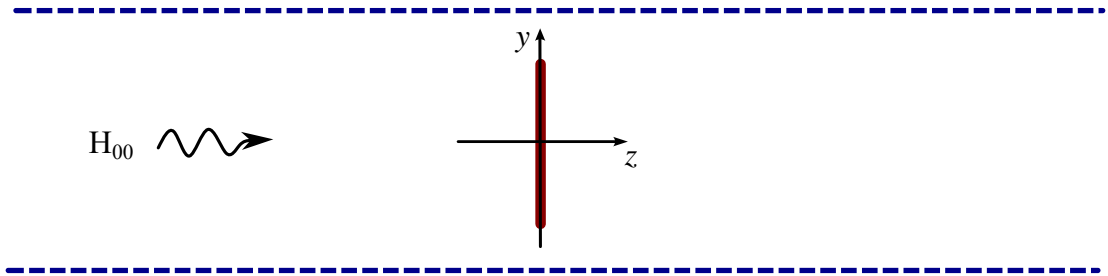

Fig. 2. Resistive film in a plane waveguide with magnetic walls.

Similarly, it can be shown that for normal incidence, the IE (29) coincides with the IE for the problem of diffraction of the fundamental wave $E_{00}$ of the E-type on a symmetric resistive film in a plane waveguide with electric walls. The solution to this problem was obtained in works [19, 20].

## 3. Vector integral equation for the tangential electric field

The purpose of this section is to formulate the IE for the tangential electric field on the interval $|y| \leq \frac{l}{2}$. For this we introduce the notation:

$$E_x(y,0) = g_x(y) e^{ik_y y},$$
$$E_y(y,0) = g_y(y) e^{ik_y y}. \tag{38}$$

Substituting $E_x(y,0)$ from (38) into the expansion (10), we obtain

$$g_x(y) = E_0 + \sum_{n=-\infty}^{\infty} a_n e^{i\frac{2\pi n}{l} y}, \tag{39}$$

whence, according to Fourier's theorem

$$a_n + \delta_{n0} E_0 = \frac{1}{2l} \int_{-\frac{l}{2}}^{\frac{l}{2}} dy \, g_x(y), \quad n = 0, \pm 1, \pm 2, \ldots. \tag{40}$$

Substituting (38) into the expansion (12), we find

$$\delta_{n0} H_0 - A_n = B_n = \frac{(k_0^2 - k_x^2)}{2k_0 l \gamma_n} \eta_0 \int_{-\frac{l}{2}}^{\frac{l}{2}} dy \, g_y(y) e^{-i\frac{2\pi n}{l} y} +$$
$$+ \frac{k_x h_n}{2k_0 l \gamma_n} \eta_0 \int_{-\frac{l}{2}}^{\frac{l}{2}} dy \, g_x(y) e^{-i\frac{2\pi n}{l} y}. \tag{41}$$

Let us now express the components of the magnetic field in terms of the tangential electric field on the periodic structure:

$$H_x(y,z) = 2H_0 e^{-ik_z z} e^{ik_y y} \sigma(z) -$$
$$- \text{sign}(z) e^{ik_y y} \int_{-\frac{l}{2}}^{\frac{l}{2}} dy' \left[ \sum_{n=-\infty}^{\infty} \eta_0 \frac{k_x h_n}{2k_0 l \gamma_n} e^{i\frac{2\pi n}{l}(y-y')} e^{i\gamma_n |z|} \right] g_x(y') -$$
$$- \text{sign}(z) e^{ik_y y} \int_{-\frac{l}{2}}^{\frac{l}{2}} dy' \left[ \sum_{n=-\infty}^{\infty} \eta_0 \frac{k_0^2 - k_x^2}{2k_0 l \gamma_n} e^{i\frac{2\pi n}{l}(y-y')} e^{i\gamma_n |z|} \right] g_y(y'), \tag{42a}$$

$$H_y(y,z) = -\frac{2}{k_0^2 - k_x^2}\left[\eta_0 E_0 k_0 k_z + k_x k_y H_0\right]e^{-ik_z z}e^{ik_y y}\sigma(z) +$$

$$+\text{sign}(z)e^{ik_y y}\int_{-\frac{l}{2}}^{\frac{l}{2}} dy' \left[\sum_{n=-\infty}^{\infty} \eta_0 \frac{k_0^2 - h_n^2}{2k_0 l \gamma_n} e^{i\frac{2\pi n}{l}(y-y')} e^{i\gamma_n |z|}\right] g_x(y') + \tag{42b}$$

$$+\text{sign}(z)e^{ik_y y}\int_{-\frac{l}{2}}^{\frac{l}{2}} dy' \left[\sum_{n=-\infty}^{\infty} \eta_0 \frac{k_x h_n}{2k_0 l \gamma_n} e^{i\frac{2\pi n}{l}(y-y')} e^{i\gamma_n |z|}\right] g_y(y'),$$

where

$$\sigma(z) = \begin{cases} 1, & z \geq 0, \\ 0, & z < 0, \end{cases}$$
$$\text{sign}(z) = \begin{cases} 1, & z \geq 0, \\ -1, & z < 0. \end{cases} \tag{43}$$

Substituting expansions (42) into boundary conditions (5b), we obtain a vector IE for determining the tangential electric field on the interval $|y| \leq \frac{l}{2}$:

$$\int_{-\frac{l}{2}}^{\frac{l}{2}} dy' \eta_{xx}(y, y') g_x(y') + \int_{-\frac{l}{2}}^{\frac{l}{2}} dy' \eta_{xy}(y, y') g_y(y') + Y(y) g_x(y) =$$
$$= -\frac{k_0 k_z}{k_0^2 - k_x^2}\eta_0 E_0 - \frac{k_x k_y}{k_0^2 - k_x^2} H_0, \quad |y| \leq \frac{l}{2}, \tag{44}$$
$$\int_{-\frac{l}{2}}^{\frac{l}{2}} dy' \eta_{yx}(y, y') g_x(y') + \int_{-\frac{l}{2}}^{\frac{l}{2}} dy' \eta_{yy}(y, y') g_y(y') + Y(y) g_y(y) =$$
$$= H_0, \quad |y| \leq \frac{l}{2}.$$

Here

$$Y(y) = \begin{cases} \dfrac{1}{W}, & |y| \leq \dfrac{d'}{2}, \\ 0, & \dfrac{d'}{2} < |y| \leq \dfrac{l}{2}, \end{cases} \tag{45}$$

while the components of the admittance tensor $\boldsymbol{\eta}$ are given by

$$\eta_{xx}(y, y') = \frac{\eta_0}{2k_0 l} \sum_{n=-\infty}^{\infty} \frac{k_0^2 - h_n^2}{\gamma_n} e^{i\frac{2\pi n}{l}(y-y')},$$

$$\eta_{xy}(y, y') = \frac{\eta_0 k_x}{2k_0 l} \sum_{n=-\infty}^{\infty} \frac{h_n}{\gamma_n} e^{i\frac{2\pi n}{l}(y-y')},$$

$$\eta_{yx}(y, y') = \eta_{xy}(y, y'),$$

$$\eta_{yy}(y, y') = \frac{\eta_0 (k_0^2 - k_x^2)}{2k_0 l} \sum_{n=-\infty}^{\infty} \frac{1}{\gamma_n} e^{i\frac{2\pi n}{l}(y-y')}.$$

(46)

Similarly to how it was done in paragraph 2, it can be shown that in the case of oblique ($k_x = 0$) or arbitrary ($k_x \neq 0$) incidence of the wave on the grating of perfectly conducting strips, the system of integral equations (44) splits into two independent IEs for E- and H-polarizations.

## 4. Determination of the difference between the tangential components of the electromagnetic field of the fundamental wave on a periodic grating

As follows from the results of the previous sections, a periodic grating can be considered as a semi-transparent film that reflects a portion of the energy, transmits another portion, and converts some of the energy into different types of waves. Let us determine the difference between the tangential components of the electric and magnetic fields of the fundamental wave on the grating. From formulas (10), (12), taking into account (14), we find for the tangential components of the electric field of the fundamental wave

$$E_x\big|_{z=+0} - E_x\big|_{z=-0} = 0,$$
$$E_y\big|_{z=+0} - E_y\big|_{z=-0} = 0,$$

(47)

$$E_x\big|_{z=0} = (E_0 + a_0) e^{ik_y y},$$

(48a)

$$E_y\big|_{z=0} = \frac{-e^{ik_y y}}{k_0^2 - k_x^2} \left[ k_x k_y (E_0 + a_0) + k_0 k_z \zeta_0 (-H_0 + A_0) \right].$$

(48b)

From formulas (11) and (13) we find the jump of the tangential components of the magnetic field on the grating

$$H_x\big|_{z=+0} - H_x\big|_{z=-0} = 2A_0 e^{ik_y y},$$

(49a)

$$H_y\big|_{z=+0} - H_y\big|_{z=-0} = \frac{2e^{ik_y y}}{k_0^2 - k_x^2} (\eta_0 k_0 k_z a_0 - k_x k_y A_0).$$

(49b)

Substituting (48a) into equation (48b) we obtain

$$(k_0^2 - k_x^2) E_y\big|_{z=0} + k_x k_y E_x\big|_{z=0} = -\zeta_0 k_0 k_z (-H_0 + A_0).$$

(50)

For the field of the fundamental wave, whose dependence on the coordinates $x$ and $y$ is specified by the factors $\exp(ik_x x)$ and $\exp(ik_y y)$, the following relations are valid:

$$\frac{\partial}{\partial x} = ik_x, \quad \frac{\partial}{\partial y} = ik_y.$$

(51)

Taking into account Maxwell's equation

$$\frac{\partial}{\partial x}E_y - \frac{\partial}{\partial y}E_x = ik_0\zeta_0 H_z \tag{52}$$

and relations (51), we transform the left side of equality (50)

$$\left(k_0^2 - k_x^2\right)E_y\big|_{z=0} + k_x k_y E_x\big|_{z=0} = k_0^2 E_y\big|_{z=0} + \left(\frac{\partial^2}{\partial x^2}E_y - \frac{\partial}{\partial x}\frac{\partial}{\partial y}E_x\right)\bigg|_{z=0} =$$
$$= k_0^2 E_y\big|_{z=0} + ik_0\zeta_0 \frac{\partial H_z}{\partial x}\bigg|_{z=0}. \tag{53}$$

Let us substitute (53) into formula (50). Then we get

$$k_0^2 E_y\big|_{z=0} + ik_0\zeta_0 \frac{\partial}{\partial x}H_z\bigg|_{z=0} = -\zeta_0 k_0 k_z e^{ik_y y}\left(-H_0 + A_0\right). \tag{54}$$

From (54) we obtain

$$e^{ik_y y} = \frac{1}{(H_0 - A_0)}\frac{k_0}{k_z}\left[\eta_0 E_y\big|_{z=0} + \frac{i}{k_0}\frac{\partial}{\partial x}H_z\bigg|_{z=0}\right]. \tag{55}$$

Substituting (55) into (49a), we get

$$H_x\big|_{z=+0} - H_x\big|_{z=-0} = \frac{2A_0}{H_0 - A_0}\frac{k_0}{k_z}\left[\eta_0 E_y\big|_{z=0} + \frac{i}{k_0}\frac{\partial}{\partial x}H_z\bigg|_{z=0}\right]. \tag{56}$$

Expression (56) gives us the sought for value for the jump of the $H_x$-component of the magnetic field in the plane $z=0$. To obtain the corresponding expression for the jump of the $H_y$ component, we substitute equation (49a) into (49b):

$$H_y\big|_{z=+0} - H_y\big|_{z=-0} = 2\eta_0 a_0 \frac{k_0 k_z}{k_0^2 - k_x^2}e^{ik_y y} - \frac{k_x k_y}{k_0^2 - k_x^2}\left(H_x\big|_{z=+0} - H_x\big|_{z=-0}\right), \tag{57}$$

whence, taking into account (51), we find

$$\left(k_0^2 + \frac{\partial^2}{\partial x^2}\right)\left[H_y\big|_{z=+0} - H_y\big|_{z=-0}\right] - \frac{\partial}{\partial x}\frac{\partial}{\partial y}\left[H_x\big|_{z=+0} - H_x\big|_{z=-0}\right] = 2a_0\eta_0 k_0 k_z e^{ik_y y}. \tag{58}$$

Taking into account Maxwell's equation

$$\frac{\partial}{\partial x}H_y - \frac{\partial}{\partial y}H_x = -ik_0\eta_0 E_z, \tag{59}$$

we transform the left side of equality (58):

$$\left(k_0^2 + \frac{\partial^2}{\partial x^2}\right)\left[H_y\big|_{z=+0} - H_y\big|_{z=-0}\right] - \frac{\partial}{\partial x}\frac{\partial}{\partial y}\left[H_x\big|_{z=+0} - H_x\big|_{z=-0}\right] =$$
$$= k_0^2\left[H_y\big|_{z=+0} - H_y\big|_{z=-0}\right] - ik_0\eta_0 \frac{\partial}{\partial x}\left[E_z\big|_{z=+0} - E_z\big|_{z=-0}\right]. \tag{60}$$

Substituting (60) into (58) we obtain:

$$2a_0\eta_0 k_0 k_z e^{ik_y y} = k_0^2 \left[ H_y\big|_{z=+0} - H_y\big|_{z=-0} \right] - ik_0 \eta_0 \frac{\partial}{\partial x}\left[ E_z\big|_{z=+0} - E_z\big|_{z=-0} \right]. \tag{61}$$

As follows from (48a)

$$e^{ik_y y} = \frac{E_x\big|_{z=0}}{E_0 + a_0}. \tag{62}$$

From (61) and (62) we find

$$E_x\big|_{z=0} = \frac{E_0 + a_0}{2a_0}\frac{k_0}{k_z}\left[ \zeta_0\left(H_y\big|_{z=+0} - H_y\big|_{z=-0}\right) - \frac{i}{k_0}\frac{\partial}{\partial x}\left(E_z\big|_{z=+0} - E_z\big|_{z=-0}\right) \right]. \tag{63}$$

Expressions (56) and (63) provide the basis for formulating equivalent boundary conditions for the electromagnetic field on grating of resistive strips. To obtain the final expressions, it is necessary to determine the quantities $\dfrac{A_0}{H_0 - A_0}$ and $\dfrac{E_0 + a_0}{a_0}$.

## 5. Determination of the quantity $\dfrac{E_0 + a_0}{a_0}$

Using the IE (22) and excitation formulas (19), we construct a variational expression of the Schwinger type [21] for the reflection coefficient $a_0$ of the fundamental wave of E-polarization, which is stationary with respect to small variations in the surface current density on the resistive film near the exact solution of the IE (22):

$$a_0 = -\zeta_0 \frac{k_0^2 - k_x^2}{2lk_0 k_z}\left[(h_y,1) + \frac{k_x k_y}{k_0^2 - k_x^2}(h_x,1)\right]\left\{\begin{array}{l} E_0\left[(h_y,1) + \dfrac{k_x k_y}{k_0^2 - k_x^2}(h_x,1)\right] - \\ -\zeta_0 \dfrac{k_0 k_z}{k_0^2 - k_x^2} H_0(h_x,1) \end{array}\right\} \times \frac{1}{T_e}, \tag{64a}$$

$$T_e = \langle h_x | \zeta_{xx} | h_x \rangle + \langle h_x | \zeta_{xy} | h_y \rangle + \langle h_y | \zeta_{yx} | h_x \rangle + \langle h_y | \zeta_{yy} | h_y \rangle + W(h_x, h_x) + W(h_y, h_y), \tag{64b}$$

where

$$(g(y), u(y)) = \int_{-\frac{d'}{2}}^{\frac{d'}{2}} dy\, g(y) u(y), \quad \langle h_m | \zeta_{mn} | h_n \rangle = \int_{-\frac{d'}{2}}^{\frac{d'}{2}} dy\, h_m(y) \int_{-\frac{d'}{2}}^{\frac{d'}{2}} dy'\, \zeta_{mn}(y, y') h_n(y'). \tag{64c}$$

We will limit ourselves to obtaining the boundary conditions in the approximation $W \ll \zeta_0$. In this case, we can assume that $h_x \ll h_y$. Therefore, let us set $h_x = 0$. Then we obtain

$$\frac{a_0}{E_0} = -\frac{\zeta_0 \dfrac{k_0^2 - k_x^2}{2k_0 l k_z}(h_y,1)^2}{\langle h_y | \zeta_{yy} | h_y \rangle + W(h_y, h_y)}. \tag{65a}$$

By isolating in $\zeta_{yy}$ the term related to the value $n = 0$, we represent the reflection coefficient $a_0$ in the following form:

$$\frac{a_0}{E_0} = -\left[1 + \sum_{n=-\infty}^{\infty \oplus} \frac{k_z}{\gamma_n} \frac{\langle h_y | e^{i\frac{2\pi n}{l}(y-y')} | h_y \rangle}{(h_y,1)^2} + \frac{W}{\zeta_0} \frac{2k_0 l k_z}{k_0^2 - k_x^2} \frac{(h_y, h_y)}{(h_y,1)^2}\right]^{-1}. \tag{65b}$$

Here the sign $\oplus$ means that the term with $n = 0$ is excluded from the sum. Comparing the obtained stationary expression for $a_0$ with the corresponding expression

$$\frac{a_0}{E_0} = -\frac{1}{1 + 2Z_e}, \tag{66}$$

found from the equivalent circuit (Fig. 3)

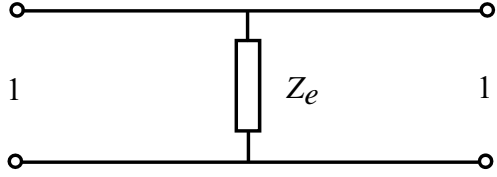

Fig. 3. The equivalent circuit.

we obtain

$$Z_e = Z_0 + \frac{W}{\zeta_0} \frac{k_0 l k_z}{k_0^2 - k_x^2} \frac{(h_y, h_y)}{(h_y,1)^2},$$

$$Z_0 = \frac{1}{2} \sum_{n=-\infty}^{\infty \oplus} \frac{k_z}{\gamma_n} \frac{\langle h_y | e^{i\frac{2\pi n}{l}(y-y')} | h_y \rangle}{(h_y,1)^2}. \tag{67}$$

Expression (63) includes the factor $\dfrac{E_0 + a_0}{2a}$, for which using (66) one obtains

$$\frac{E_0 + a_0}{2a_0} = -Z_e. \tag{68}$$

We substitute expression (68) into condition (63) and transform it taking into account (51) and (60). Then we get

$$E_x|_{z=0} = -Z_0 \frac{k_0}{k_z} \left[\zeta_0 \left(H_y|_{z=+0} - H_y|_{z=-0}\right) - \frac{i}{k_0}\frac{\partial}{\partial x}\left(E_z|_{z=+0} - E_z|_{z=-0}\right)\right] - \\ -Wl \frac{(h_y,h_y)}{(h_y,1)^2}\left(H_y|_{z=+0} - H_y|_{z=-0}\right) - Wl \frac{k_x k_y}{k_0^2 - k_x^2}\frac{(h_y,h_y)}{(h_y,1)^2}\left(H_x|_{z=+0} - H_x|_{z=-0}\right). \tag{69}$$

For $W \ll \zeta_0$ the following inequality holds

$$\left(H_x|_{z=+0} - H_x|_{z=-0}\right) \ll \left(H_y|_{z=+0} - H_y|_{z=-0}\right) \tag{70a}$$

and the last term in (69) can be neglected. Then we get

$$E_x\big|_{z=0} = -Z_0 \frac{k_0}{k_z}\left[\zeta_0\left(H_y\big|_{z=+0} - H_y\big|_{z=-0}\right) - \frac{i}{k_0}\frac{\partial}{\partial x}\left(E_z\big|_{z=+0} - E_z\big|_{z=-0}\right)\right] - $$
$$- Wl\frac{(h_y, h_y)}{(h_y,1)^2}\left(H_y\big|_{z=+0} - H_y\big|_{z=-0}\right). \tag{70b}$$

## 6. Determination of the quantity $\dfrac{A_0}{H_0 - A_0}$

Using the IE (44) and excitation formulas (41), we construct a variational expression of the Schwinger type [21] for the reflection coefficient $A_0$ of the fundamental wave of H-polarization, which is stationary with respect to small variations of the tangential electric field on the interval $|y| \leq \dfrac{l}{2}$:

$$H_0 - A_0 = \frac{(k_0^2 - k_x^2)}{2k_0 l k_z}\eta_0\left[(g_y,1) + \frac{k_x k_y}{k_0^2 - k_x^2}(g_x,1)\right] \times$$
$$\times\left\{H_0\left[(g_y,1) - \frac{k_x k_y}{k_0^2 - k_x^2}(g_x,1)\right] - \frac{k_0 k_z}{k_0^2 - k_x^2}\eta_0(g_x,1)\right\} \times \frac{1}{T_h}, \tag{71}$$
$$T_h = \langle g_x|\eta_{xx}|g_x\rangle + \langle g_x|\eta_{xy}|g_y\rangle + \langle g_y|\eta_{yx}|g_x\rangle + \langle g_y|\eta_{yy}|g_y\rangle + (Yg_x, g_x) + (Yg_y, g_y).$$

When $W \ll \zeta_0$ we can assume that $g_x \ll g_y$. Therefore, as a first approximation, we set $g_x = 0$. Then we get

$$1 - \frac{A_0}{H_0} = \frac{\eta_0 \dfrac{k_0^2 - k_x^2}{2k_0 l k_z}(g_y,1)^2}{\langle g_y|\eta_{yy}|g_y\rangle + (Yg_y, g_y)}. \tag{72}$$

By selecting the term corresponding to $n=0$ in the expression $\eta_{yy}$ from (72), we represent $\left(1 - \dfrac{A_0}{H_0}\right)$ in the following form

$$1 - \frac{A_0}{H_0} = \left[1 + \sum_{n=-\infty}^{\infty \oplus} \frac{k_z}{\gamma_n}\frac{\langle g_y|e^{i\frac{2\pi n}{l}(y-y')}|g_y\rangle}{(g_y,1)^2} + \zeta_0 \frac{2k_0 l k_z}{k_0^2 - k_x^2}\frac{(Yg_y, g_y)}{(g_y,1)^2}\right]^{-1}. \tag{73}$$

Comparing the obtained stationary expression (73) with the expression

$$1 - \frac{A_0}{H_0} = \left[1 + \frac{Y_h}{2}\right]^{-1}, \tag{74}$$

following from the equivalent circuit (Fig.3), where the conductivity of the parallel element is designated $Y_h$, we obtain

$$Y_h = Y_0 + \Delta Y,$$

$$Y_0 = 2 \sum_{n=-\infty}^{\infty \oplus} \frac{k_z}{\gamma_n} \frac{\left\langle g_y \left| e^{i\frac{2\pi n}{l}(y-y')} \right| g_y \right\rangle}{(g_y, 1)^2}, \qquad (75)$$

$$\Delta Y = \zeta_0 \frac{4 l k_0 k_z}{k_0^2 - k_x^2} \frac{(Yg_y, g_y)}{(g_y, 1)^2}.$$

In expression (56) for the jump of the $H_x$ component of the magnetic field on the grating, the factor $\dfrac{2A_0}{H_0 - A_0}$ is included. As follows from (74), it is equal to $Y_h$. Let us substitute $Y_h$ from (75) into condition (56):

$$H_x\big|_{z=+0} - H_x\big|_{z=-0} = Y_0 \frac{k_0}{k_z} \left[ \eta_0 E_y\big|_{z=0} + \frac{i}{k_0} \frac{\partial H_z}{\partial x}\bigg|_{z=0} \right] + \\ + \Delta Y \frac{k_0}{k_z} \left[ \eta_0 E_y\big|_{z=0} + \frac{i}{k_0} \frac{\partial H_z}{\partial x}\bigg|_{z=0} \right]. \qquad (76a)$$

We transform the second term on the right-hand side of (76a) taking into account expression (53). Then we obtain

$$H_x\big|_{z=+0} - H_x\big|_{z=-0} = Y_0 \frac{k_0}{k_z} \left[ \eta_0 E_y\big|_{z=0} + \frac{i}{k_0} \frac{\partial H_z}{\partial x}\bigg|_{z=0} \right] + \\ + 4l \frac{(Yg_y, g_y)}{(g_y, 1)^2} E_y\big|_{z=0} + 4l \frac{k_x k_y}{k_0^2 - k_x^2} \frac{(Yg_y, g_y)}{(g_y, 1)^2} E_x\big|_{z=0}. \qquad (76b)$$

When $\dfrac{W}{\zeta_0} \ll 1$ the last term in (76b) can be neglected. Then we have

$$H_x\big|_{z=+0} - H_x\big|_{z=-0} = Y_0 \frac{k_0}{k_z} \left[ \eta_0 E_y\big|_{z=0} + \frac{i}{k_0} \frac{\partial H_z}{\partial x}\bigg|_{z=0} \right] + \\ + 4l \frac{(Yg_y, g_y)}{(g_y, 1)^2} E_y\big|_{z=0}. \qquad (77)$$

## 7. Calculation of the impedance of the shunt $Z_e$ of the equivalent circuit in the problem of diffraction of the fundamental E-polarized wave

Let us now proceed to the calculation of the impedance $Z_e$ of the parallel element of the equivalent circuit in the problem of diffraction of the fundamental E-polarized wave on a grating of resistive strips. As shown in works [19, 20], the magnetic field jump component $h_y$, which is transverse to the strip, behaves as $O(\rho^0)$ ($\rho$ is the distance to the edge of the strip) when approaching the edge. Therefore, we define the trial function for the value $h_y(y)$ in the form

$$h_y(y) = \text{const}, \quad |y| \leq \frac{d'}{2}. \tag{78}$$

Substituting (78) into the expression (67), we find

$$\frac{(h_y, h_y)}{(h_y, 1)^2} = (d')^{-1}. \tag{79}$$

For $k_0 l \ll 1$ and $n \neq 0$ we have $\gamma_n = i\frac{2\pi |n|}{l}$. Therefore, we get

$$Z_0 \frac{k_0}{k_z} = -i\frac{l}{2\lambda} \sum_{n=-\infty}^{\infty \oplus} \frac{\langle 1 | e^{i\frac{2\pi n}{l}(y-y')} | 1 \rangle}{|n|(d')^2}. \tag{80}$$

Integrating in (80) with respect to $y$ and $y'$

$$\int_{-\frac{d'}{2}}^{\frac{d'}{2}} dy\, e^{i\frac{2\pi n}{l}y} = \int_{-\frac{d'}{2}}^{\frac{d'}{2}} dy\, e^{-i\frac{2\pi n}{l}y} = 2\int_0^{\frac{d'}{2}} dy \cos\left(\frac{2\pi n}{l} y\right) =$$
$$= \frac{l}{\pi n} \sin\left(\frac{\pi n d'}{l}\right), \tag{81}$$

we obtain

$$Z_0 \frac{k_0}{k_z} = -i\frac{l}{\lambda}\left(\frac{l}{\pi d'}\right)^2 \sum_{n=1}^{\infty} \frac{\sin^3\left(\frac{\pi n d'}{l}\right)}{n^3}. \tag{82}$$

For $\frac{d'}{l} \ll 1$ we have [22]

$$\sum_{n=1}^{\infty} \frac{\sin^2\left(\frac{\pi n d'}{l}\right)}{n^3} = \left(\frac{\pi d'}{l}\right)^2 \left[\ln\left(\frac{l}{2\pi d'}\right) + \frac{3}{2}\right]. \tag{83a}$$

Then for $\frac{d'}{l} \ll 1$ we obtain

$$Z_0 \frac{k_0}{k_z} = -i\frac{l}{\lambda}\left[\ln\left(\frac{l}{2\pi d'}\right) + \frac{3}{2}\right]. \tag{84a}$$

For $\frac{d}{l} \ll 1$ we have

$$\sum_{n=1}^{\infty} \frac{\sin^2\left(\frac{\pi n d'}{l}\right)}{n^3} = \sum_{n=1}^{\infty} \frac{\sin^2\left(\frac{\pi n d}{l}\right)}{n^3} = \left(\frac{\pi d}{l}\right)^2 \left[\ln\left(\frac{l}{2\pi d}\right) + \frac{3}{2}\right] \tag{83b}$$

and

$$Z_0 \frac{k_0}{k_z} = -i\frac{l}{\lambda}\left(\frac{d}{l}\right)^2\left[\ln\left(\frac{l}{2\pi d}\right)+\frac{3}{2}\right]. \tag{84b}$$

Substituting now (79) and (82) into formula (70), we obtain the boundary condition

$$E_x\big|_{z=0} = i\frac{l}{\lambda}\left(\frac{l}{\pi d'}\right)^2 \sum_{n=1}^{\infty} \frac{\sin^2\left(\frac{\pi n d'}{l}\right)}{n^3}\left[\zeta_0\left(H_y\big|_{z=+0} - H_y\big|_{z=-0}\right) - \frac{i}{k_0}\frac{\partial}{\partial x}\left(E_z\big|_{z=+0} - E_z\big|_{z=-0}\right)\right] - $$
$$-W\frac{l}{d'}\left(H_y\big|_{z=+0} - H_y\big|_{z=-0}\right). \tag{85}$$

## 8. Calculation of the admittance $Y_h$ of the shunt of the equivalent circuit in the problem of diffraction of the fundamental H-polarized wave

Let us now proceed to the calculation of the admittance $Y_h$ of the parallel element of the equivalent circuit in the problem of diffraction of the fundamental H-polarized wave on a grating of resistive strips, which is determined by formula (75). The most difficult part is choosing the trial distribution of the y-component of the electric field, which must be specified both on the strip and on the slits. Since we limited ourselves to the case of $\frac{W}{\zeta_0} \ll 1$, then in calculating $Y_0$ we can assume that $g_y$ coincides with the distribution of the y-component of the electric field in the problem of diffraction on a grating of perfectly conducting strips. In this case, it is shown in work [21] that in the static approximation $\frac{l}{\lambda} \ll 1$

$$g_y(y) = \begin{cases} 0, & |y| \leq \frac{d'}{2} \\ \dfrac{\sin\left(\dfrac{\pi y}{l}\right)}{\sqrt{\left(\sin\dfrac{\pi d}{2l}\right)^2 - \left(\cos\dfrac{\pi y}{l}\right)^2}}, & \dfrac{d'}{2} \leq |y| \leq \dfrac{l}{2}. \end{cases} \tag{86}$$

$$Y_0 \frac{k_0}{k_z} = i\frac{4l}{\lambda}\ln\left[\cos\left(\frac{\pi d'}{2l}\right)\right]. \tag{87}$$

However, in this approximation we will get a zero value for the quantity $\Delta Y$, since outside the strip $Y = 0$. To select the trial function $g_y$ we will use the results of the study of the distribution of the electromagnetic field near the edge of the resistive half-plane [19, 20]. It is shown there that function $g_y$ behaves differently depending on whether one is approaching the edge of the resistive half-plane (Fig. 4) from the free space or along the resistive half-plane. Hereinafter we omit the dimensional factor in function $g_y$.

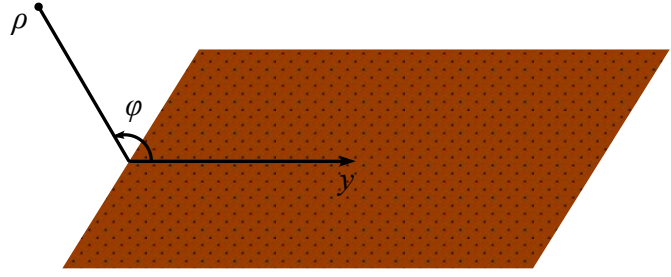

Fig. 4. Resistive half-plane.

When approaching from the free space, the $g_y$-component becomes infinite according to the law

$$g_y\big|_{\varphi=\pi} = \frac{A}{\sqrt{\rho}}, \qquad (88)$$

while when approaching along the resistive half-plane $g_y$ goes to zero according to the square root law

$$g_y\big|_{\varphi=0} = B\sqrt{\rho}. \qquad (89)$$

As follows from the results of works [19, 20], the ratio $\dfrac{B}{A}$ is equal to

$$\frac{B}{A} = 4\frac{W}{\zeta_0} k_0. \qquad (90)$$

Functions satisfying conditions (88)-(90) can be defined in the following form:

<u>on the slits</u> $\left(\dfrac{d'}{2} \leq |y| \leq \dfrac{l}{2}\right)$

$$g_y(y) = \frac{1}{\sqrt{\left(\dfrac{d}{2}\right)^2 - u^2}}, \quad u = \frac{l}{2} - |y|. \qquad (91)$$

<u>on the resistive strip</u> $\left(|y| \leq \dfrac{d'}{2}\right)$

$$g_y(y) = 4k_0 \frac{W}{\zeta_0} \sqrt{\frac{d}{d'}} \sqrt{\left(\frac{d'}{2}\right)^2 - y^2}. \qquad (92)$$

Let us calculate the arising integrals:

$$(Yg_y, g_y) = \frac{W}{\zeta_0} \eta_0 \frac{16k_0^2}{d'd} \int_{-\frac{d'}{2}}^{\frac{d'}{2}} dy \left[\left(\frac{d'}{2}\right)^2 - y^2\right] = \frac{W}{\zeta_0} \eta_0 \frac{8k_0^2}{3d} (d')^2. \qquad (93)$$

$$\left(g_y,1\right) = \int_{-\frac{d'}{2}}^{\frac{d'}{2}} du \frac{1}{\sqrt{\left(\frac{d}{2}\right)^2 - u^2}} = \pi. \tag{94}$$

Then

$$\frac{\left(Yg_y, g_y\right)}{\left(g_y,1\right)^2} = \frac{W}{\zeta_0}\eta_0 \frac{32}{3d}\left(\frac{d'}{\lambda}\right)^2. \tag{95}$$

Substituting (87) and (95) into expression (77) we obtain the boundary condition

$$H_x\big|_{z=+0} - H_x\big|_{z=-0} = i\frac{4l}{\lambda}\ln\left[\cos\left(\frac{\pi d'}{2l}\right)\right]\left[\eta_0 E_y\big|_{z=0} + \frac{i}{k_0}\frac{\partial H_z}{\partial x}\bigg|_{z=0}\right] + \\ + \frac{W}{\zeta_0}\frac{l}{d}\frac{128}{3}\left(\frac{d'}{\lambda}\right)^2 \eta_0 E_y\big|_{z=0}. \tag{96}$$

## Conclusions

In this work, we have obtained boundary conditions that allow one to consider a periodic grating of resistive strips as a plane of discontinuity of the electromagnetic field. These conditions can be written in the following form:

$$E_x\big|_{z=+0} - E_x\big|_{z=-0} = 0,$$
$$E_y\big|_{z=+0} - E_y\big|_{z=-0} = 0,$$

$$E_x\big|_{z=0} = i\frac{l}{\lambda}\left(\frac{l}{\pi d'}\right)^2 \sum_{n=1}^{\infty} \frac{\sin^2\left(\frac{\pi n d'}{l}\right)}{n^3}\left[\zeta_0\left(H_y\big|_{z=+0} - H_y\big|_{z=-0}\right) - \frac{i}{k_0}\frac{\partial}{\partial x}\left(E_z\big|_{z=+0} - E_z\big|_{z=-0}\right)\right] - \\ - W\frac{l}{d'}\left(H_y\big|_{z=+0} - H_y\big|_{z=-0}\right).$$

$$H_x\big|_{z=+0} - H_x\big|_{z=-0} = i\frac{4l}{\lambda}\ln\left[\cos\left(\frac{\pi d'}{2l}\right)\right]\left[\eta_0 E_y\big|_{z=0} + \frac{i}{k_0}\frac{\partial H_z}{\partial x}\bigg|_{z=0}\right] + \\ + \frac{W}{\zeta_0}\frac{l}{d}\frac{128}{3}\left(\frac{d'}{\lambda}\right)^2 \eta_0 E_y\big|_{z=0}.$$